\documentclass[conference]{IEEEtran}
\IEEEoverridecommandlockouts
\usepackage{cite}
\usepackage{amsmath,amssymb,amsfonts, dsfont, booktabs}
\usepackage{algorithmic}
\usepackage{graphicx}
\usepackage{textcomp, hyperref}
\usepackage{xcolor}

\hypersetup{
    colorlinks=true,
    linkcolor=black,
    filecolor=magenta,      
    urlcolor=magenta,
    pdftitle={Overleaf Example},
    pdfpagemode=FullScreen,
    }

\def\BibTeX{{\rm B\kern-.05em{\sc i\kern-.025em b}\kern-.08em
    T\kern-.1667em\lower.7ex\hbox{E}\kern-.125emX}}
    
\begin{document}

\title{Introducing Self-Attention to Target Attentive Graph Neural Networks
}

\author{\IEEEauthorblockN{1\textsuperscript{st} Sai Mitheran}
\IEEEauthorblockA{\textit{National Institute of Technology,}\\
\textit{Tiruchirappalli}\\
saimitheran06@gmail.com}
\and
\IEEEauthorblockN{2\textsuperscript{nd} Abhinav Java}
\IEEEauthorblockA{\textit{Delhi Technological University} \\
java.abhinav99@gmail.com}
\and
\IEEEauthorblockN{3\textsuperscript{rd} Surya Kant Sahu}
\IEEEauthorblockA{\textit{The Learning Machines} \\
surya.oju@pm.me}
\and
\IEEEauthorblockN{4\textsuperscript{th} Arshad Shaikh}
\IEEEauthorblockA{\textit{BYJU's} \\
arshaikh5775@gmail.com}
}

\maketitle

\begin{abstract}
Session-based recommendation systems suggest relevant items to users by modeling user behavior and preferences using short-term anonymous sessions. Existing methods leverage Graph Neural Networks (GNNs) that propagate and aggregate information from neighboring nodes i.e., local message passing. Such graph-based architectures have representational limits, as a single sub-graph is susceptible to overfit the sequential dependencies instead of accounting for complex transitions between items in different sessions. We propose a new technique that leverages a Transformer in combination with a target attentive GNN. This allows richer representations to be learnt, which translates to empirical performance gains in comparison to a vanilla target attentive GNN. Our experimental results and ablation show that our proposed method is competitive with the existing methods on real-world benchmark datasets, improving on graph-based hypotheses. Code is available at \url{https://github.com/The-Learning-Machines/SBR}.
\end{abstract}

\begin{IEEEkeywords}
Session-based recommendation, Graph neural networks, Transformers
\end{IEEEkeywords}

\section{Introduction}
Traditional recommendation systems use user-item interactions across multiple sessions to model user preferences. However, in session-based recommendation (SBR) systems, the users are anonymous; hence inter-session data cannot be used. The goal here is to predict the items with which the user is likely to interact, given previous item interactions within a single session. In general, the number of user-item interactions is limited, and as a result, modeling user intent is a challenging task. Nevertheless, session-based recommendation is gaining momentum due to increasing privacy concerns. Recent advancements in session-based recommendation systems have focused on modeling the user-item interaction as a directed graph and hence leverages graph-based architectures and related multi-level feature extraction techniques. However, these methods are disposed to representational limits \cite{garg2020generalization} as a sub-graph tends to overfit the sequential dependencies, while the essence of extracting the representations that connote complex transitions between multi-session items are lost.

\begin{figure}[!t]
    \centering
    \includegraphics[width=0.7\linewidth]{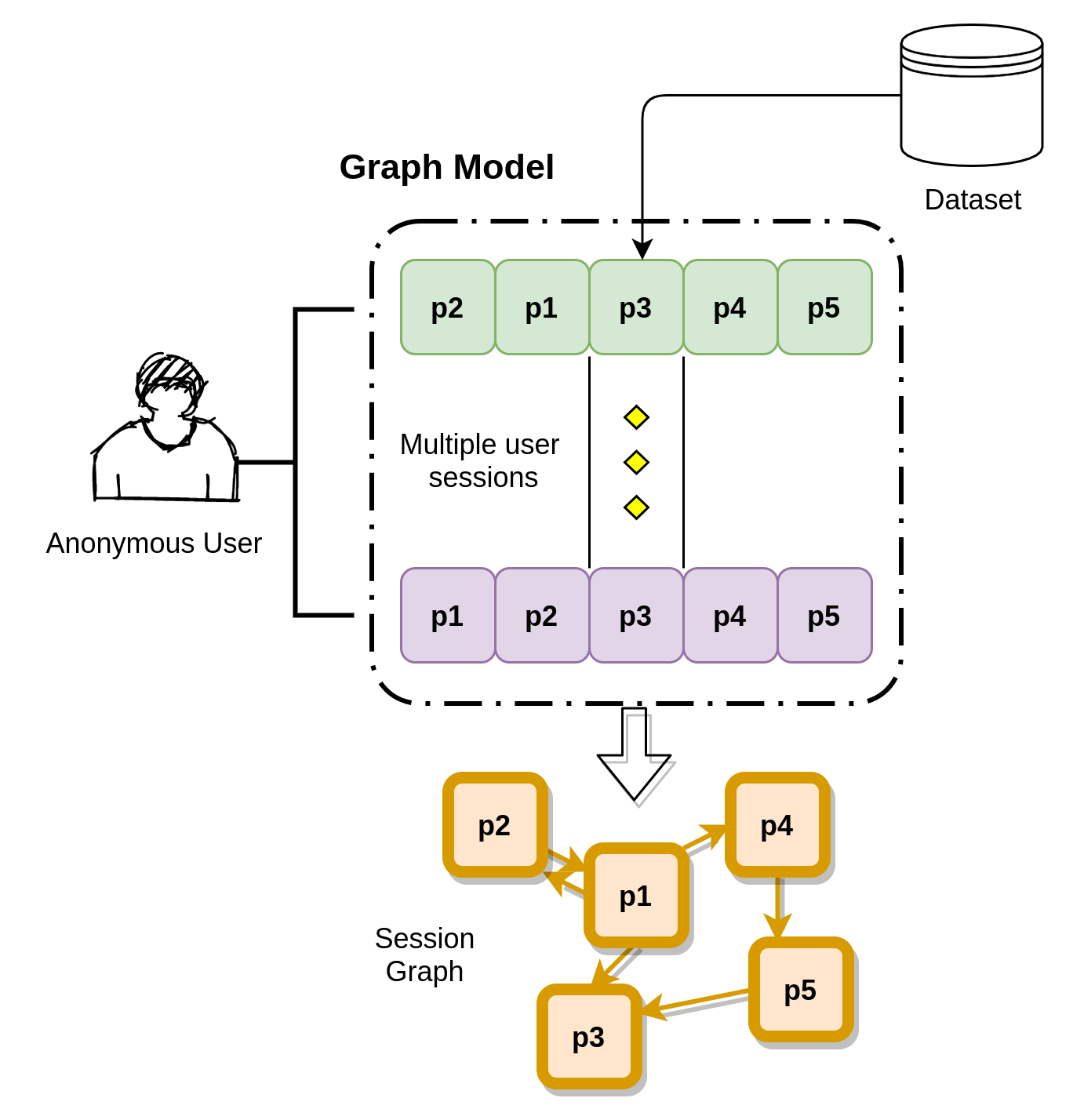}
    \caption{Session-based Recommendation Systems}
    \label{task}
\end{figure}

In this work we present TAGNN++, incorporating Transformers as universal function approximators to enhance capturing complex transitions that address the limitations of GNNs in learning rich representations. We model item interactions using a GNN and both global and local user interaction with a Transformer. We show that our model is competitive with the existing state-of-the-art techniques on the Diginetica benchmark and on the Yoochoose benchmark. We also explore Adaptive Gradient Clipping \cite{brock2021high} for Transformer-based architectures specific to our task and present an ablation study to analyze the performance of the model.

% In this work we present TAGNN++, incorporating Transformers as universal function approximators to enhance capturing complex transitions that address the limitations of GNNs in learning rich representations. We model item interactions using a GNN and both global and local user interaction with a Transformer. We show that our model is competitive with the existing state-of-the-art techniques on the Diginetica benchmark\footnote{http://cikm2016.cs.iupui.edu/cikm-cup} and on the Yoochoose benchmark\footnote{http://2015.recsyschallenge.com/challenge.html}. We also explore Adaptive Gradient Clipping \cite{brock2021high} for Transformer-based architectures specific to our task and present an ablation study to analyze the performance of the model.

\section{Related works}

\subsection{Deep Learning based SBRs}

% A widely used technique to learn the user-item interaction function is through GNNs. The simple GNNs can also be combined with GRUs [], Convolution Layers [], and Attention mechanisms [] to leverage individual advantages. Each session can be mapped into a graph’s chain. Each item is represented as a node and the interactions as edges. The natural compatibility between data modeled in such a manner and the GNNs allows this method to perform well and was first introduced in \cite{Wu:2019ke}.

GNNs offer an intuitive approach to session-based recommendation since each session can be mapped into a graph's chain. Each node of the graph represents an item, and each edge represents an interaction. The natural compatibility between data modeled in such a manner and GNNs allow this method to perform well and was introduced in \cite{Wu:2019ke}. Despite the convenient representation of sessions offered by GNNs, it lacks the ability to model long-range dependencies and intricate interactions as substantiated in Graph-Contextualised Self-Attention model \cite{ijcai2019-547}. Qiu \textit{et al.} \cite{Rethinking} proposed using a weighted graph attention layer for focusing on the essential parts of the item embeddings. Extending the idea of using attention with GNNs, Yu \textit{et al.}\cite{yu2020tagnn} proposed using both item embeddings with a GNN and target embedding with an attention mechanism to achieve significant performance gain. However, a GNN applied on a single sub-graph is susceptible to overfit on sequential dependencies instead of accounting for complex transitions between items in different sessions. Using a dual-channel GNN capable of complex item transition modeling addresses this issue \cite{dgtn}. 
% We restrict our comparison to this approach since it instills a dependency on neighboring session graphs, which has questionable scaling and deployment in real-time scenarios for rigorous practical applications in session-based recommendation systems.

% Introducing attention to this framework was seen in the Graph-Contextualised Self-Attention model \cite{ijcai2019-547} to account for long-range dependencies.

% Further, we can focus on item embeddings using a Weighted Graph Attention layer proposed in \cite{Rethinking}. However, recent studies of shifting attention to the target item have shown to give increased performance \cite{yu2020tagnn}. This target-attentive graph neural network first learns embeddings for items utilizing a GNN and then uses an attention mechanism to activate different user interests according to varied target items for the task at hand. This tends to bias over sequential dependencies since we apply a GNN on a single sub-graph, not accounting for multi-session information. Here, there is information loss as there is slack on weighing global dependencies. We need to learn complex inter-session global dependencies and improve the methods to handle extracted features. \textcolor{red}{DGTN to be added and format as mentioned}

\subsection{Use of Adaptive Gradient Clipping (AGC)}

Clipping the gradient is a commonly used approach to improving gradient descent \cite{Zhang2020WhyGC}, but manual selection of the clipping threshold increases the number of hyperparameters. However, the hyperparameter for gradient clipping needs to be tuned carefully, as it is susceptible to the loss function and architecture \cite{9231926}. The chosen threshold is vital as if it is set too high, then the gradient norm will always be smaller than that, and clipping is never applied. If too low, then the network's step size may be too small and cause convergence issues leading to unstable learning. We can clip gradients based on the unit-wise ratio of gradient norms to parameter norms \cite{brock2021high}. This helps ensure stable training across different batch sizes, allowing larger learning rates to ensure quick convergence, and is effective to tackle poorly conditioned loss landscapes. By incorporating this, we verify that performance is marginally increased for Transformer architectures specific to our task.

\section{Method}

\subsection{Problem Formulation and Preliminaries}
The SBR problem contains three types of entities : Users, Items, and User-item interactions. Here, we formally introduce the SBR problem and associated notations. Let $S = [S_0, S_1, S_2, ..., S_{N-1}]$ be the set of $N$ sessions and $V = [V_0, V_1, V_2, ..., V_{M-1}]$ be the set of $M$ unique items in the dataset. Let $S_k = [x_0, x_1, x_2, ..., x_t]$ be the item sequence within a session $S_k$, where $ x_t \in V$ is the item clicked by a user at time step $t = 0, 1, 2 ... $. The goal is to estimate the parameter set $\theta$ such that $p(x_{t+1}|x_0, x_1, x_2, ..., x_t; \theta)$ is the maximum-likelihood estimator of $x_{t+1}$, where $k \in [0, N-1]$.

The proposed TAGNN++ leverages two key components, namely - \emph{Graph Neural Networks and Self-Attention}. We use the same background strategy for constructing the session graph from the given data as proposed in \cite{yu2020tagnn} and \cite{Wu:2019ke}. Each session is stored in memory as an adjacency matrix representing a directed graph. In this graph $\mathds{G_s}$, the $i^{th}$ node in the node set $\mathds{V_{s}}$ represents an item such that $n_{s, i} \in \mathds{V_{s}}$. Each edge connecting nodes $i$ and $j$ denotes subsequent item selections by the user in the given session $s$ such that $(n_{s, i}, n_{s, j}) \in \mathds{E_s}$ \cite{yu2020tagnn}. Lastly, each node has an incoming weight and an outgoing weight associated to it. We effectively capture item transitions and connections using a gated GNN \cite{li2015gated}, by learning both item and session embeddings as in \cite{yu2020tagnn}.

% add equations

\subsection{Proposed method - TAGNN++}
We observe that a simple attention model is unable to capture both local and global context \cite{NIPS2017_3f5ee243}, hence indicating the need for a more robust way to represent sequences. Furthermore, GNNs build a representation of data by message passing or neighborhood aggregation, using only local information. Garg \textit{et al.} show that even simple graph structures are indistinguishable by GNNs relying only on local information, making it hard for a GNN to compute several graph properties \cite{garg2020generalization}. Additionally, they perform a thorough study of generalization bounds for message passing, which accentuates the need for better function approximators.

\begin{figure}[htb]
    \centering
    \includegraphics[width=0.7\linewidth]{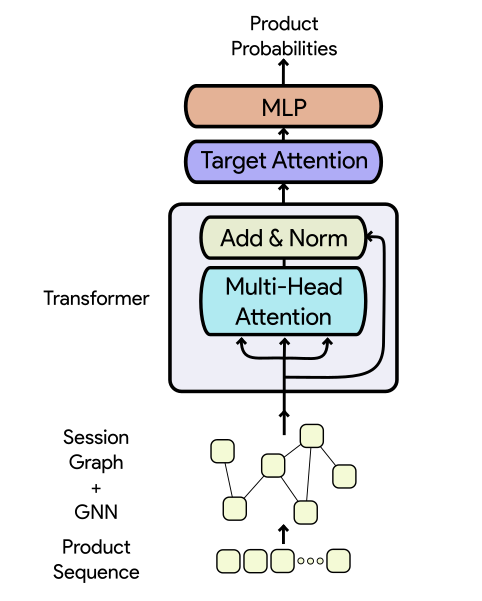}
    \caption{Proposed Architecture : TAGNN++}
    \label{arch}
\end{figure}

% \begin{figure}[htb]
%     \centering
%     \includesvg{figures/Architectures/TAGNN++.svg}
%     \caption{Proposed Architecture : TAGNN++}
%     \label{arch}
% \end{figure}

To that end, we propose a multi-headed Transformer-based design for target-aware predictions. The proposed design enables us to leverage both GNNs and powerful attention models; in essence, model item interaction in the form of a graph and a user's long and short-term interaction through the Transformer. We hypothesize that our design is a \textit{superior function approximator} for the task of next-item prediction in session-based recommendation, providing empirical evidence and a carefully crafted ablative study. Recently, Transformers have been successfully used in Natural Language Processing and Computer Vision by parallelizing the self-attention \cite{NIPS2017_3f5ee243} mechanism over multiple heads achieving state of the art results. \cite{yun2019Transformers} also gives a compact support based proof of why Transformers have high representation capacity and are \textit{universal approximators of permutation equivariant sequence to sequence functions}. 

At the $l^{th}$ layer the following update takes place in a single head:

\begin{equation}
    h_{i}^{l+1} = \sum_{\forall j \in S} \sigma(\frac{Q^{l}h_{i}^{l} \cdot (K^{l}h_{j}^{l})^{T}}{\sqrt{d_{k}}}) V^{l}
\end{equation}

where $h_{j}^{l}$ is the hidden state of the $j^{th}$ item in sequence $S$ at layer $l$ such that $h_{j}^{l} \forall j \in S$. $Q^{l}$, $K^{l}$, $V^{l}$ are Queries, Keys and Values respectively, and $\sigma$ denotes the softmax operation. $Q^{l}$, $K^{l}$, $V^{l}$ are learnable parameters that are updated in parallel as opposed to sequentially, and $d_k$ is a hyperparameter indicating the dimension of the linear layer in the Transformer. In our approach, we allow the Transformer to concatenate several representations or transformations of the interaction between the hidden states of the target and sequence. This is done using multiple such attention heads and the concatenation of the outputs before normalizing them across the layer, before being passed on to a Feed-forward module. At the feature level, it is possible that the values in the attention matrix cover a large domain, which is not desirable since the network finds it hard to learn the optimal parameter across different scales quickly. This is where Layer Normalization comes into play by normalizing across the feature space.

Finally, the token-wise Feed-forward Layer (FFL) transforms the normalized context vector to the output sequence. The composition of these FFLs implicitly implement a scalar quantization map such that each input is mapped to the output \cite{yun2019Transformers}. We stress that the proposed augmentations to the TAGNN architecture make it more robust to different kinds of data streams. The various operations in the Transformer that include Self-attention, Layer Norm, and the Feed-forward Layer with skip connections play a vital role in enabling Transformers to be universal approximators of sequence-to-sequence functions.

% The sequence of operations in the Transformer from self-attention layer, layer norm to feed forward along with skip connections play an important role in allowing being universal approximators of sequence to sequence functions.

\begin{table}[htb]
\centering
\caption{Comparison of several baseline methods with the proposed method.}
\begin{tabular}{lllll}
\toprule
\multicolumn{1}{c}{\textbf{Method}}                                            & \multicolumn{2}{c}{\textbf{Yoochoose 1/64}} & \multicolumn{2}{c}{\textbf{Diginetica}} \\ \midrule
                                                                        & HR@20            & MRR@20           & HR@20          & MRR@20         \\ \midrule
% POP                                                                & 6.71           & 1.65            & 0.89          & 0.20          \\ \hline

% Item-KNN \cite{item-knn}                                                                & 51.60            & 21.81            & 35.75          & 1.57           \\ \hline
% BPR-MF  \cite{BPR}                                                                 & 31.31            & 12.08            & 5.24           & 1.98           \\ \hline
% FPMC    \cite{FPMC}                                                                & 45.62            & 15.01            & 26.53          & 6.95           \\ 
GRU4REC  \cite{hidasi2015session}                                                               & 60.64            & 22.89            & 29.45          & 8.33           \\
NARM     \cite{li2017neural}                                                               & 68.32            & 28.63            & 49.70          & 16.17          \\
% STAMP \cite{STAMP}                                                                  & 68.74            & 29.67            & 45.64          & 14.32          \\
RepeatNet  \cite{repeatnet}                                                             & 70.71            & 31.03            & 47.79          & 17.66          \\
CSRM    \cite{CSRM}                                                                & 71.45            & 30.36            & 50.55          & 16.38          \\
SR-GNN  \cite{Wu:2019ke}                                                                & 70.87            & 30.94            & 50.73          & 17.59          \\
GC-SAN \cite{ijcai2019-547}                                                                 & 70.66            & 30.04            & 51.70          & 17.61          \\
TAGNN \cite{yu2020tagnn}                                                                & 71.02            & 31.12            & 51.31          & 18.03          \\
LESSR  \cite{chen2020handling}                                                                 & 70.64            & 30.97            & 51.71          & \textbf{18.15}          \\  \midrule
% DGTN  \cite{dgtn}                                                                 & 71.18            & 31.35            & 53.05         & 18.07          \\ \hline 
% \textbf{\begin{tabular}[c]{@{}l@{}}Transformer SBR\end{tabular}} & 71.97          & 31.62          & 52.04          & 17.62         % \\ \hline
% \\
% \textbf{\begin{tabular}[c]{@{}l@{}}Transformer\end{tabular}}     & \textbf{72.15 }         & 30.79          & 51.31        & 17.14         \\

% \textbf{\begin{tabular}[c]{@{}l@{}}TAGNN++\\ \end{tabular}} & 70.04 & {32.53} & {55.46} & {19.51} \\ %\hline
\textbf{\begin{tabular}[c]{@{}l@{}}TAGNN++\end{tabular}} & \textbf{71.91} & \textbf{31.57} & \textbf{51.86} & 17.93 \\ \bottomrule 
\end{tabular}
\label{tabcomparison}
\end{table}

\section{Setup and Experimental Results}

% In this section, we study the performance of our proposed architecture - TAGNN++, for Next-item prediction in Session-based Recommendation.

\textbf{\textit{Evaluation Metrics}}: We use two metrics from previous studies, i.e., Hit Rate@N, and MRR@N, where N $= 20$. Hit Rate calculates the number of "hits" in an N-sized list of ranked items, where a "hit" refers to something that the user has clicked on, purchased, or "saved for later", based on context. Mean reciprocal rank (MRR) is used to judge a system where the order/positions of the retrieved items are important.

\textbf{\textit{Datasets}}: For testing our hypothesis, we employ two widely used real-world datasets, Yoochoose\footnotemark[2] and Diginetica\footnotemark[1]. Table \ref{dataset} provides information on the dataset contents. We use only the recent 1/64 fraction of the Yoochoose dataset, denoted as Yoochoose 1/64, which comprises various sessions that specify the clicks by a given user. It is a collection of records in a file containing a Session ID, Timestamp, Item ID, and Category. Diginetica contains anonymized search and browsing logs, product data, anonymized transactions, and an extensive collection of product images.

\textbf{\textit{Preprocessing}}: For simplicity, we apply the same preprocessing as \cite{yu2020tagnn}, \cite{Wu:2019ke}. We drop all unit length sessions and remove items that appear less than five times for both datasets as same as previous studies. For generating training and test sets, sessions of last days are used as the test set for Yoochoose 1/64, and sessions of last weeks as the test set for Diginetica. For an existing session, we generate a series of input session sequences and corresponding labels. We filter out items from the test set which do not appear in the training set.

\textbf{\textit{Hyperparameters}}: We retain most hyperparameter settings from previous baselines to display the advantage of learning better representations using Transformers. We keep $10\%$ of our datasets for validation and use a batch size of $50$, for $15$ epochs. We use the Adam \cite{da2014method} optimizer with an initial learning rate of $10^{-4}$, with momentum parameters $\beta_{1}$ = $0.9$, $\beta_{2}$ = $0.999$, and decay it by a factor of $0.1$ every $3$ epochs. Additionally, we set the $L2$ penalty to $10^{-6}$. 
For Multi-Head Attention in the Transformer, we set the number of heads as $2, 8$ and dropout of $0.1$, with the embedding dimension as $100, 120$ for Yoochoose 1/64 and Diginetica respectively.
% For Multi-Head Attention in the Transformer, we set the number of heads as $2/8$\footnote{Format - Yoochoose/Diginetica} and dropout of $0.1$, with the embedding dimension as $100/120$\footnotemark[4]. 

\textbf{\textit{Baselines}}: We compare our method with baseline GNN and Attention-based methods for session-based recommendation, as shown in Table \ref{tabcomparison}. The RNN-based methods \cite{hidasi2015session} \cite{li2017neural} \cite{CSRM}, and further \cite{repeatnet} that takes repeat consumption into account were then outperformed by graph-based methods \cite{Wu:2019ke}, those involving the notion of attention \cite{ijcai2019-547}  \cite{yu2020tagnn}.

\begin{table}[htb]
\centering
\caption{Ablation Study}
\label{ablation}
\begin{tabular}{lllll}
\toprule
\multicolumn{1}{c}{\textbf{Architecture}}                 & \multicolumn{2}{c}{\textbf{Yoochoose 1/64}} & \multicolumn{2}{c}{\textbf{Diginetica}} \\ \midrule
     & \multicolumn{1}{c}{HR@20} & MRR@20  & HR@20   & MRR@20  \\ \midrule
\textbf{TAGNN++}    &                                        \textbf{71.91} & 31.57 & \textbf{51.86} & \textbf{17.93}   \\ 
- AGC & 71.80                    & 31.41 & 51.57 & 17.65 \\ 
- GNN & 71.75           & \textbf{31.62} & 51.48 & 17.59  \\ 
- PE  & 71.69                      & 31.48   & 51.64   & 17.66   \\ 
- Transformer & 71.03                      & 30.69   & 51.42   & 17.84 \\ \bottomrule
% \begin{tabular}[c]{@{}l@{}}Inverted \\ TAGNN++\end{tabular} & 70.04           & 32.71                      & 55.7313             & 19.414             \\ \hline
\end{tabular}
\end{table}

\begin{table}[htb]
\centering
\caption{Dataset Stats}
\begin{tabular}{llclll}
\toprule
\multicolumn{1}{c}{\textbf{Dataset}} &
  \textbf{\begin{tabular}[c]{@{}l@{}}Clicks\end{tabular}} &
  \textbf{\begin{tabular}[c]{@{}c@{}} Train\\ Sessions\end{tabular}} &
  \textbf{\begin{tabular}[c]{@{}l@{}}Test \\ Sessions\end{tabular}} &
  \textbf{\begin{tabular}[c]{@{}l@{}}Total \\ Items\end{tabular}} &
  \textbf{\begin{tabular}[c]{@{}l@{}}Avg. \\ Length\end{tabular}} \\ \midrule
Diginetica &
  982961 &
  719470 &
  68977 &
  43074 &
  5.13 \\ \hline
Yoochoose 1/64 &
  557248 &
  \multicolumn{1}{l}{375043} &
  55898 &
  16339 &
  6.11 \\ \bottomrule
\end{tabular}
\label{dataset}
\end{table}

\begin{figure}[htb]
    \centering
    \includegraphics[width=0.85\linewidth]{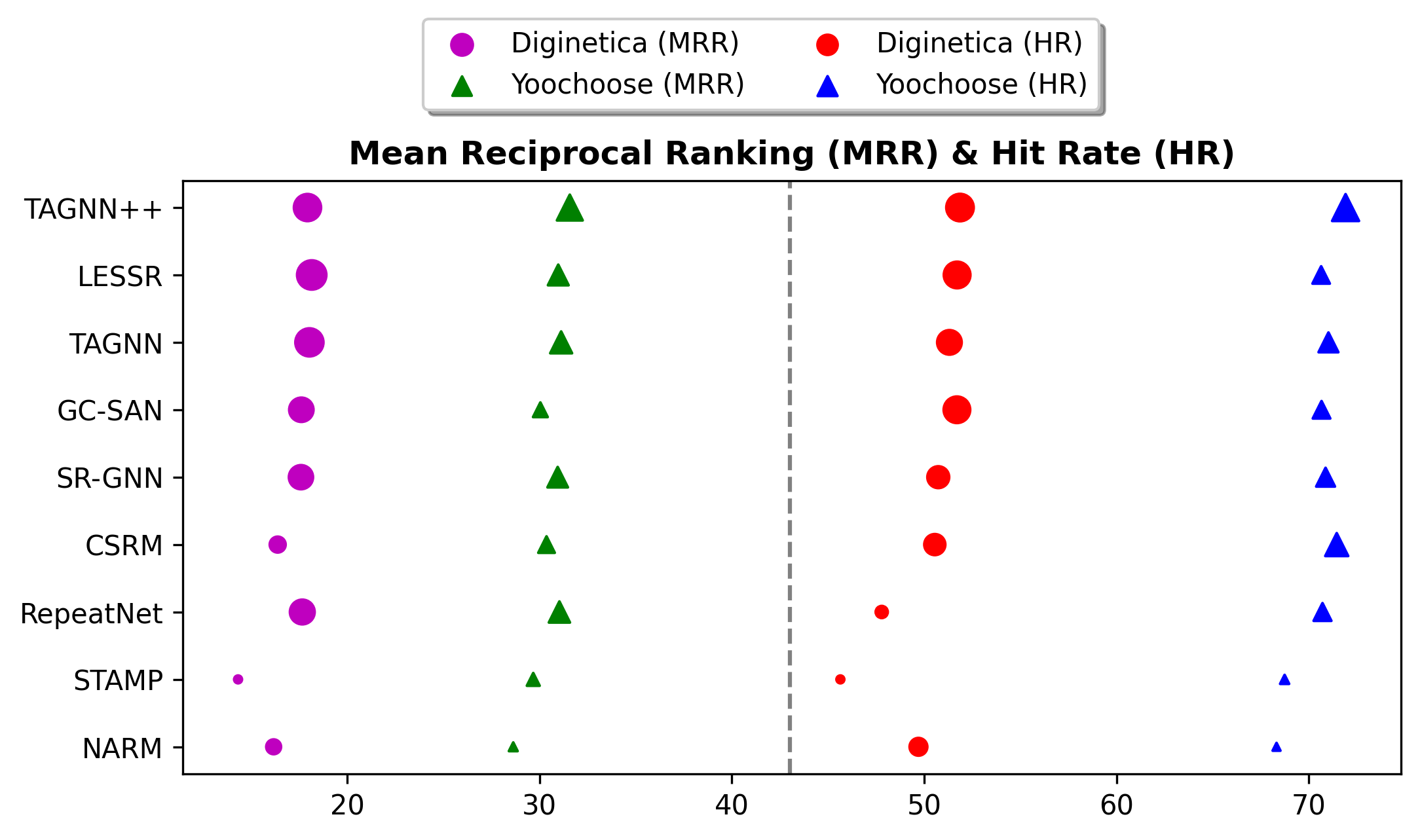}
    \caption{Scatter plot comparison of models on the evaluation metrics}
    \label{fig:eval_sbr}
\end{figure}

      \textbf{\textit{Inferences}}: We can infer from the values of HR@20 and MRR@20 that our method is competitive with the previous state-of-the-art on the given benchmarks, whilst outperforming the graph-based TAGNN model \cite{yu2020tagnn}. Hence, it is evident that Deep learning-based methods are more capable of learning a better user-item interaction representation by \emph{capturing complex data distributions and transitions} in the dataset. Incorporating Transformers to overcome the representational limitations in a GNN improves the performance of our recommendation system. The improvement observed verifies that learning better representations as proposed is helpful to model complex patterns, as shown in Fig. ~\ref{fig:eval_sbr}. To verify the efficacy of our architecture, we also perform an ablation study as shown in Table \ref{ablation}. It can be observed that the performance is only marginally affected by the removal of secondary techniques such as AGC (Adaptive Gradient Clipping) and PE (Positional Embedding). Further, the removal of the Transformer decreases the overall performance, verifying that GNNs fail at learning accurate representations in some cases.
% We also depict the importance of the Transformer block using the Inverted TAGNN++ architecture. We switch the GNN and Transformer blocks, which only slightly impacts the model's performance.

Fig. ~\ref{fig:eval_sbr} indicates better performance on Yoochoose 1/64 than Diginetica, due to a greater average session length in the former - as more historical data would enhance the Transformer’s attention mechanism to leverage global context. Both datasets consist of short, medium, and long sessions \cite{Wang2019ASO}. When the user wishes to build a session-based recommender consisting of long sessions (Yoochoose 1/64) with better rank/order as the main objective (tweets, webpages, music), we show that our model outperforms existing methods by a considerable margin. For short/medium length sessions (Diginetica), where both order and the number of desired items in the top N of the ranked list are important (shopping items, movies/videos), our model is competitive with the existing methods. 

% \begin{figure}[htb]
%     \centering
%     \includegraphics[width=0.8\linewidth]{figures/Eval/Ablation_Eval_Plot_Vertical.png}
%     \caption{Ablation Study}
%     \label{fig:abl_sbr}
% \end{figure}
% \vspace{-2em}
\section{Conclusion}
This paper proposes an alternative method for learning richer representations in session-based recommendation models, to overcome the limitations posed by message-passing GNNs. We perform intelligent feature extraction using Transformers with target attentive GNNs. We leverage Multi-head Attention to capture both local and global context. Our method is competitive with the previous state-of-the-art on the aforementioned benchmarks. We show that our method is suitable for rank-based retrieval in long sessions (Yoochoose), and also establish competitive performance for short/medium length sessions (Diginetica). Motivating research in this direction would enable making informative and practical choices in streaming, business operations, and E-Commerce.

\bibliographystyle{IEEEtran}
\bibliography{sample-base}

\begin{thebibliography}{10}
\providecommand{\url}[1]{#1}
\csname url@samestyle\endcsname
\providecommand{\newblock}{\relax}
\providecommand{\bibinfo}[2]{#2}
\providecommand{\BIBentrySTDinterwordspacing}{\spaceskip=0pt\relax}
\providecommand{\BIBentryALTinterwordstretchfactor}{4}
\providecommand{\BIBentryALTinterwordspacing}{\spaceskip=\fontdimen2\font plus
\BIBentryALTinterwordstretchfactor\fontdimen3\font minus
  \fontdimen4\font\relax}
\providecommand{\BIBforeignlanguage}[2]{{%
\expandafter\ifx\csname l@#1\endcsname\relax
\typeout{** WARNING: IEEEtran.bst: No hyphenation pattern has been}%
\typeout{** loaded for the language `#1'. Using the pattern for}%
\typeout{** the default language instead.}%
\else
\language=\csname l@#1\endcsname
\fi
#2}}
\providecommand{\BIBdecl}{\relax}
\BIBdecl

\bibitem{garg2020generalization}
V.~Garg, S.~Jegelka, and T.~Jaakkola, ``Generalization and representational
  limits of graph neural networks,'' in \emph{International Conference on
  Machine Learning}.\hskip 1em plus 0.5em minus 0.4em\relax PMLR, 2020, pp.
  3419--3430.

\bibitem{brock2021high}
A.~Brock, S.~De, S.~L. Smith, and K.~Simonyan, ``High-performance large-scale
  image recognition without normalization,'' \emph{arXiv preprint
  arXiv:2102.06171}, 2021.

\bibitem{Wu:2019ke}
\BIBentryALTinterwordspacing
S.~Wu, Y.~Tang, Y.~Zhu, L.~Wang, X.~Xie, and T.~Tan, ``Session-based
  recommendation with graph neural networks,'' in \emph{Proceedings of the
  Twenty-Third AAAI Conference on Artificial Intelligence}, P.~V. Hentenryck
  and Z.-H. Zhou, Eds., vol.~33, no.~1.\hskip 1em plus 0.5em minus 0.4em\relax
  AAAI Press, Jul. 2019, pp. 346--353. [Online]. Available:
  \url{https://aaai.org/ojs/index.php/AAAI/article/view/3804}
\BIBentrySTDinterwordspacing

\bibitem{ijcai2019-547}
\BIBentryALTinterwordspacing
C.~Xu, P.~Zhao, Y.~Liu, V.~S. Sheng, J.~Xu, F.~Zhuang, J.~Fang, and X.~Zhou,
  ``Graph contextualized self-attention network for session-based
  recommendation,'' in \emph{Proceedings of the Twenty-Eighth International
  Joint Conference on Artificial Intelligence, {IJCAI-19}}.\hskip 1em plus
  0.5em minus 0.4em\relax International Joint Conferences on Artificial
  Intelligence Organization, 7 2019, pp. 3940--3946. [Online]. Available:
  \url{https://doi.org/10.24963/ijcai.2019/547}
\BIBentrySTDinterwordspacing

\bibitem{Rethinking}
\BIBentryALTinterwordspacing
R.~Qiu, J.~Li, Z.~Huang, and H.~YIn, ``Rethinking the item order in
  session-based recommendation with graph neural networks,'' in
  \emph{Proceedings of the 28th ACM International Conference on Information and
  Knowledge Management}, ser. CIKM '19.\hskip 1em plus 0.5em minus 0.4em\relax
  New York, NY, USA: Association for Computing Machinery, 2019, p. 579–588.
  [Online]. Available: \url{https://doi.org/10.1145/3357384.3358010}
\BIBentrySTDinterwordspacing

\bibitem{yu2020tagnn}
F.~Yu, Y.~Zhu, Q.~Liu, S.~Wu, L.~Wang, and T.~Tan, ``Tagnn: Target attentive
  graph neural networks for session-based recommendation,'' in
  \emph{Proceedings of the 43rd International ACM SIGIR Conference on Research
  and Development in Information Retrieval}, 2020, pp. 1921--1924.

\bibitem{dgtn}
Y.~{Zheng}, S.~{Liu}, Z.~{Li}, and S.~{Wu}, ``Dgtn: Dual-channel graph
  transition network for session-based recommendation,'' in \emph{2020
  International Conference on Data Mining Workshops (ICDMW)}, 2020, pp.
  236--242.

\bibitem{Zhang2020WhyGC}
J.~Zhang, T.~He, S.~Sra, and A.~Jadbabaie, ``Why gradient clipping accelerates
  training: A theoretical justification for adaptivity,'' \emph{arXiv:
  Optimization and Control}, 2020.

\bibitem{9231926}
P.~Seetharaman, G.~Wichern, B.~Pardo, and J.~L. Roux, ``Autoclip: Adaptive
  gradient clipping for source separation networks,'' in \emph{2020 IEEE 30th
  International Workshop on Machine Learning for Signal Processing (MLSP)},
  2020, pp. 1--6.

\bibitem{li2015gated}
Y.~Li, D.~Tarlow, M.~Brockschmidt, and R.~Zemel, ``Gated graph sequence neural
  networks,'' \emph{arXiv preprint arXiv:1511.05493}, 2015.

\bibitem{NIPS2017_3f5ee243}
\BIBentryALTinterwordspacing
A.~Vaswani, N.~Shazeer, N.~Parmar, J.~Uszkoreit, L.~Jones, A.~N. Gomez, L.~u.
  Kaiser, and I.~Polosukhin, ``Attention is all you need,'' in \emph{Advances
  in Neural Information Processing Systems}, I.~Guyon, U.~V. Luxburg,
  S.~Bengio, H.~Wallach, R.~Fergus, S.~Vishwanathan, and R.~Garnett, Eds.,
  vol.~30.\hskip 1em plus 0.5em minus 0.4em\relax Curran Associates, Inc.,
  2017. [Online]. Available:
  \url{https://proceedings.neurips.cc/paper/2017/file/3f5ee243547dee91fbd053c1c4a845aa-Paper.pdf}
\BIBentrySTDinterwordspacing

\bibitem{yun2019Transformers}
C.~Yun, S.~Bhojanapalli, A.~S. Rawat, S.~J. Reddi, and S.~Kumar, ``Are
  transformers universal approximators of sequence-to-sequence functions?''
  \emph{arXiv preprint arXiv:1912.10077}, 2019.

\bibitem{hidasi2015session}
B.~Hidasi, A.~Karatzoglou, L.~Baltrunas, and D.~Tikk, ``Session-based
  recommendations with recurrent neural networks,'' \emph{arXiv preprint
  arXiv:1511.06939}, 2015.

\bibitem{li2017neural}
J.~Li, P.~Ren, Z.~Chen, Z.~Ren, T.~Lian, and J.~Ma, ``Neural attentive
  session-based recommendation,'' in \emph{Proceedings of the 2017 ACM on
  Conference on Information and Knowledge Management}, 2017, pp. 1419--1428.

\bibitem{repeatnet}
\BIBentryALTinterwordspacing
P.~Ren, Z.~Chen, J.~Li, Z.~Ren, J.~Ma, and M.~de~Rijke, ``Repeatnet: A repeat
  aware neural recommendation machine for session-based recommendation,''
  \emph{Proceedings of the AAAI Conference on Artificial Intelligence},
  vol.~33, no.~01, pp. 4806--4813, Jul. 2019. [Online]. Available:
  \url{https://ojs.aaai.org/index.php/AAAI/article/view/4408}
\BIBentrySTDinterwordspacing

\bibitem{CSRM}
\BIBentryALTinterwordspacing
M.~Wang, P.~Ren, L.~Mei, Z.~Chen, J.~Ma, and M.~de~Rijke, ``A collaborative
  session-based recommendation approach with parallel memory modules,'' in
  \emph{Proceedings of the 42nd International ACM SIGIR Conference on Research
  and Development in Information Retrieval}, ser. SIGIR'19.\hskip 1em plus
  0.5em minus 0.4em\relax New York, NY, USA: Association for Computing
  Machinery, 2019, p. 345–354. [Online]. Available:
  \url{https://doi.org/10.1145/3331184.3331210}
\BIBentrySTDinterwordspacing

\bibitem{chen2020handling}
T.~Chen and R.~C.-W. Wong, ``Handling information loss of graph neural networks
  for session-based recommendation,'' in \emph{Proceedings of the 26th ACM
  SIGKDD International Conference on Knowledge Discovery \& Data Mining}, 2020,
  pp. 1172--1180.

\bibitem{da2014method}
K.~Da, ``A method for stochastic optimization,'' \emph{arXiv preprint
  arXiv:1412.6980}, 2014.

\bibitem{Wang2019ASO}
S.~Wang, L.~Cao, and Y.~Wang, ``A survey on session-based recommender
  systems,'' \emph{ArXiv}, vol. abs/1902.04864, 2019.

\end{thebibliography}

\end{document}